\documentclass[copyright]{eptcs}
\providecommand{\event}{FLACOS 2012} 
\usepackage{breakurl}             

\usepackage{listings}
\usepackage{amsmath}
\usepackage{prooftree}
\usepackage{graphicx}
\usepackage{wrapfig}
\usepackage{subfigure}
\usepackage{arrows}
\usepackage{paralist}
\usepackage[strings]{underscore}

\title{A History of BlockingQueues}
\author{Marina Zaharieva-Stojanovski \qquad Marieke Huisman \qquad Stefan Blom
\institute  {University of Twente, The Netherlands
\thanks{
This work was supported by ERC grant 258405 for the VerCors project
(all authors), and Artemis grant
2008-100039 for the CHARTER project (Blom). }\\
}
\email{m.zaharieva/m.huisman/s.blom@utwente.nl}
}
\def\titlerunning{A History of BlockingQueues}
\def\authorrunning{M. Zaharieva-Stojanovski, M. Huisman, S. Blom}
\begin{document}
\maketitle


\begin{abstract}

This paper describes a way to formally specify the behaviour of concurrent data structures. When specifying concurrent data structures, the main challenge is to make specifications stable, i.e., to ensure that they cannot be invalidated by other threads. To this end, we propose to use history-based specifications: instead of describing method behaviour in terms of the object's state, we specify it in terms of the object's state history. A history is defined as a list of state updates, which at all points can be related to the actual object's state.

We illustrate the approach on the \texttt{BlockingQueue} hierarchy from
the \texttt{java.util.concurrent} library. We show how the behaviour of the interface \texttt{BlockingQueue} is specified,
leaving a few decisions open to descendant classes. 
The classes implementing the interface correctly inherit the
specifications. As a specification language, we use a combination of JML
and permission-based separation logic, including abstract predicates.
This results in an abstract, modular and natural way to specify the
behaviour of concurrent queues. The specifications can be used to derive
high-level properties about queues, for example to show that the order
of elements is preserved. Moreover, the approach can be
easily adapted to other concurrent data structures.

\end{abstract}

\section{Introduction}

Writing 'good' formal specifications is an essential factor for efficient program verification. Moreover, these specifications, written in the form of contracts of the software components, provide useful documentation with precisely defined requirements. They should be elegant, readable, independent from the code implementation, and should fully express the required code behaviour.

However, specifying concurrent software is not straightforward. Currently, different techniques are being developed to reason about concurrent programs and specific concurrency features~\cite{OwickiG75,SchellhornTER11,Haack76303}, but specifying the functional behaviour of such a program is still a challenge. To address this challenge, this paper introduces a history-based approach to develop behavioural specifications for concurrent data structures. We present the idea on a collection of realistic Java classes, using a combination of JML (the \emph{Java Modeling Language})~\cite{LeavensPCCRCMKC07} and \emph{separation logic}~\cite{OHearnRY01,Reynolds02separationlogic} as a specification language. This results in clear, readable and abstract specifications. 

The major difficulty in specifying the behaviour of concurrent code is caused by thread interference. If threads operate on the same object, a method's postcondition can be \emph{unstable}, i.e., it can be invalidated by other threads. The following code fragment is part of the \texttt{put(E e)} method from a shared \texttt{Queue} data structure. The method enqueues an element to the end of the list that internally represents the queue:
\begin{lstlisting}[language=java, basicstyle=\footnotesize]
public void put(E e){
    lock.lock(); 
    enqueue(e); 
    lock.unlock();   . . .  
}
\end{lstlisting} 
A straightforward way to define the postcondition of this method would be to state that \texttt{e} is the last element in the queue. However, after the lock release another thread may interfere, remove \texttt{e}, and invalidate the postcondition before the caller continues. Therefore, this postcondition would be unstable.

Instead, we propose to maintain a \emph{history} that records all elements that have been added to the queue. In this way, we can reason about the past. For example, for the \texttt{put} method we can use a history to state that: \emph{There was a moment in the past when \texttt{e} was added to the queue}. 

The main contributions of this work are formal history-based specifications for the classes from the \texttt{java.util.concurrent.BlockingQueue} hierarchy. To make the specifications abstract, modular and naturally organized, we use abstract predicates, specification inheritance and JML model variables. This way we provide an efficient template for easily specifying the classes that implement the \texttt{BlockingQueue} interface, even if they maintain a different order on the elements. Furthermore, we claim that the history pattern is also useful for specifying other concurrent data structures; in particular we have also specified the \texttt{ConcurrentHashMap} using the same idea. Support to reason about history-based specifications will be integrated into the tool set that is currently being developed as part of the VerCors project~\cite{AmighiBHZ12}  (Verification of Concurrent Data Structures).

\section{Functional Behaviour Specifications}

In this section we explain how we use histories to overcome the problem of \emph{unstable postconditions}. 
Mainly, we discuss the specifications for the \texttt{BlockingQueue} interface, and then we shortly describe how these specifications are inherited by the concrete \texttt{BlockingQueue} instantiation classes. 

\paragraph{History-based Specifications}

A history in our approach is a linked list of nodes that keep the references of all queue elements. The history is defined as a specification ghost variable:
\texttt{/*@ public ghost JMLValueSequence history = new JMLValueSequence();\@*/}. Therefore, each queue element \texttt{e} has an appropriate history node \texttt{e'} that holds the reference \texttt{e}. We say that \texttt{e'} is \emph{a matching node} of \texttt{e}. History updates are done by adding JML \emph{set statements} at the locations where the queue is changed (i.e., an element is added or removed).
Since the history should remember all \emph{'old'} elements, when a queue element is deleted, its matching history node is not removed, but marked with a negative flag. We define this flag in the node class as a \texttt{boolean exists} variable.
The nodes in the history are represented by a model class \texttt{ListElem<E>}, where \texttt{E} is the type of the \texttt{BlockingQueue} items:
\begin{lstlisting}[language=java , basicstyle=\footnotesize]
/*@ model class ListElem<E> implements JMLType {
  @   boolean exist = true;  
  @   E item; 
  @   int orderNum = 0; 
  @   public int compareOrderTo(ListElem<E> o){
  @       return orderNum - o.orderNum; 
  @   }    . . .    
  @ } @*/
\end{lstlisting}

\paragraph{The Order of the Elements} Our approach requires preserving the same order of the elements in the history and the queue. Therefore, it is mandatory that each node is added to the correct position in the history list. Nevertheless, different \texttt{BlockingQueue} implementations follow a different order. To keep our approach abstract, we add a \texttt{compareOrderTo(ListElem<E>)} method in the \texttt{ListElem} class, making the history nodes comparable to each other with respect to their own specific order. 
We use this method to add a new node to the history at the proper position.

The common order of the elements in a \texttt{BlockingQueue} is FIFO. Thus, we define a default implementation of the \texttt{compareOrderTo} method with respect to FIFO. Since this order (as well as LIFO) depends on the moment when an element is added to the history, we add the \texttt{orderNum} variable in the \texttt{ListElem} class to represent a counter for the element insertion to the history. Hence, the \texttt{ListElem} is a default history node that respects the FIFO order. The \texttt{BlockingQueue} implementations that follow different ordering of the elements should extend this class and override the \texttt{compareOrderTo} method. An example is the \texttt{PriorityBlockingQueue} where the order of the element depends on the item value.

\paragraph{Methods' Contracts}
We define the methods' postconditions to express the behaviour of the queue in terms of its history. 
The postcondition of the \texttt{add(E e)} method states that the history contains the matching node of \texttt{e}. We cannot express anything about the value of its \texttt{exist} flag because it is possible that another thread has removed the element before the end of the method:
\begin{lstlisting}[language=java , basicstyle=\footnotesize]
/*@ public behavior
  @ requires e != null;
  @ ensures (\exists ListElem n; history.has(n) * n.item == e); @*/
void put(E e) throws InterruptedException;
\end{lstlisting}
The postcondition of the \texttt{take()} method states that the matching node of the \texttt{\char92result} (the object returned as a result of this method) exists in the history with a \texttt{false} value of the \texttt{exist} flag. It is not sure that the same node existed before method start, but if it existed, its flag was \texttt{true}. 
\begin{lstlisting}[language=java , basicstyle=\footnotesize]
/*@ public behavior
   @ requires true;
   @ ensures \exists ListElem n; history.has(n) * n.item == \result * !n.exist *
   @ (\old(history).contains(n) ==> \old(n).exist); @*/
E take() throws InterruptedException;
\end{lstlisting}

\paragraph{Class Invariant}

\begin{wrapfigure}{r}{0.5\textwidth}
\vspace*{-0.5cm}
\centering
\includegraphics[scale=0.5]{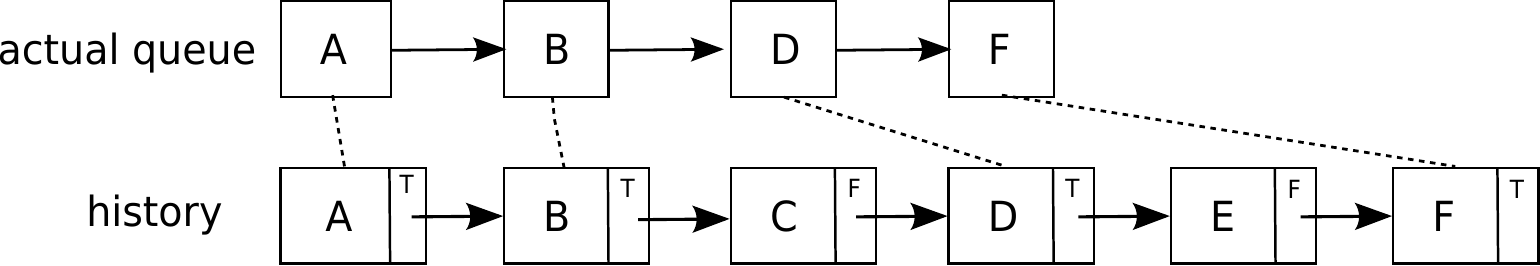}
\caption{Compatibility between the history and the queue}
\label {fig: comp}
\end{wrapfigure}

The crucial part of the specifications is to define a property to state that the queue is \emph{compatible} with the history. This ensures that the queue behaves correctly. Informally, we define a \emph{compatibility} property stating that: if all nodes with a \texttt{false} flag are removed from the history, the remaining nodes \emph{match} with the queue elements and are ordered in the same manner (see Fig.~\ref{fig: comp}).

To express this property, we define a recursive predicate \texttt{compatible}, where both parameters (\texttt{queue} and \texttt{history}) are of \texttt{JMLValueSequence} type. We could also have chosen a different type for the \texttt{queue} variable (for example, an array), but we found that this was a simple and elegant way to express the predicate. 

With the approach presented so far, the order of the history nodes \texttt{e1} and \texttt{e2} is not precisely defined if \texttt{e1.compareOrderTo(e2)=0} (in this case we call \texttt{e1} and \texttt{e2} \emph{equally ordered nodes}). Note that equally ordered nodes can exist for example in the \texttt{PriorityBlockingQueue} history. If the nodes \texttt{A} and \texttt{B} in the history (Fig.~\ref{fig: comp}) were \emph{equally ordered}, we cannot guarantee that \texttt{A} precedes or follows \texttt{B}. Anyhow, the queue and the history should be compatible and therefore, they should follow the same order. The following formal specification of the \texttt{compatible} predicate deals with this case:

\begin{lstlisting}[language=java, basicstyle=\footnotesize]
/*@ pred compatible(JMLValueSequence queue, JMLValueSequence history) = 
  @ (queue.isEmpty() && history.isEmpty()) ||                              //case 1               
  @ (!queue.isEmpty() &&
  @  (\exists int idx; 0<=idx<history.length;
  @    (((ListElem)queue.first()).item == ((ListElem)history.itemAt(idx)).item) &&
  @    ((ListElem)history.itemAt(idx)).exist &&
  @    (\forall int i; 0<=i<idx; 
  @     !((ListElem)history.itemAt(i)).exist || 
  @     (((ListElem)history.itemAt(i)).compareOrderTo((ListElem)history.itemAt(idx)) 
  @	    == 0))      
  @    && compatible(queue.removeItemAt(0), history.removeItemAt(idx))))   //case 2
  @*/
\end{lstlisting}

With this definition we even allow expressing the compatibility between queue and a history where the order of elements is not important at all (a set data structure for example). In that case, the method \texttt{compareOrderTo} should be overridden to return always 0, and the \texttt{compatible} predicate will express that the queue contains the correct elements without considering their order.

Furthermore, a class invariant is specified to state that the queue is compatible with the history. To state the invariant, we use a model variable
\texttt{queue} to denote the current contents of the queue.

\begin{lstlisting}[language=java , basicstyle=\footnotesize]
/*@ protected model instance JMLValueSequence queue;
  @ invariant compatible(queue,  history) @*/
\end{lstlisting}

\paragraph{Specifications for the Descendant Classes} 

Specifying the descendant classes is simple: the following steps are required:
 \begin{inparaenum}[(\itshape i\upshape)]
\item if the order of the elements in the queue is different then FIFO, extend the \texttt{ListElem} class and override the \texttt{compareOrderTo(ListElem<E>)} method;
\item update the \texttt{history} ghost variable at the locations where the queue is changed i.e., an element is added or removed).
\end{inparaenum}

The complete specifications together with the Java source code are available via \url{http://www.ewi.utwente.nl/~zaharievam/specs_classes/}.

\section{Conclusions}
This paper presents a history-based approach for specifying the functional behaviour of concurrent data structures. It allows one to write \emph{stable} method specifications, i.e., specifications
that cannot be invalidated by other threads.  We apply this approach on the \texttt{java.util.concurrent.BlockingQueue} hierarchy, but it is also suitable for specifying other data structures. Furthermore, our approach introduces a new dimension: abstraction. The result is a specification framework, where specifying a new \texttt{BlockingQueue} implementation requires only a small effort.

\paragraph{Related Work}

Using a history for specifying communication-based parallel programs dates back to the 70's and 80's. A history, also called a \emph{communication trace}, is a sequence that records the communication events sent between the processes. The history helps one to reason about the events that happened up to a certain point in time and their order. This approach has been used to specify the interactions in CSP~\cite{Soundararajan83}. Also in recent work, a similar idea has been used: Dovland \emph{et al.}~\cite{Dovland:2008} use histories to record the communication events between the components in a distributed environment.

The specifications we provide are closely related to the linearizability-based approach where proving that a data structure is linearizable guarantees its correctness. In this approach the specification contains sequential code that updates a ghost variable, which is an abstract representation of the data structure itself. A library is linearizable with respect to a specification if it behaves as if it was sequentially executed. Linearizability is automatically proved by searching a \emph{linearization point} for each method ~\cite{Vafeiadis10}.
 Gotsman \emph{et al.}~\cite{linown11} explain the importance of linearizability stating that: when proving properties for a client that uses a concurrent library, this library can be soundly replaced by its abstract implementation.

Our work is more practically oriented. We are focused on making clear and understandable specifications, written in a standard form of method contracts and class invariants. The specifications are readable even without the code, and therefore can be used as formally written documentation. 

\paragraph{Future Work}

The work described here are only the first steps in the VerCors
project~\cite{AmighiBHZ12}, and there is still much work to be done in
the future. A first important step is to develop tool support, so that
the history-based specifications can be validated w.r.t. the API
reference implementation. 

Another point we are currently working on is providing full support for class invariants in a concurrent environment. 
These class invariants must be \emph{strong}, i.e., invariants that hold in all possible states. 
We allow an invariant to explicitly break under certain conditions, while guaranteeing that no other thread is able to observe that it is broken.

It still has to be investigated how the history-based specifications
can be exploited to verify client's code. One particular case where
this will be useful is if the behaviour of the client depends on the
elements that have been in the queue at some point. For example, in a
game application, the queue could hold the \(k\) highest scores in
descending order. A client can use the history-based specifications to
derive properties as: the queue holds the best scores, and it knows
the correct number of total scores.

Finally, we will also investigate whether we can automatically
generate updates to the history, instead of depending on the user
manually inserting these updates in the descendant classes.

\bibliographystyle{eptcs}
\bibliography{vercors}
\end{document}